\documentclass[aps,prd,twocolumn,showpacs,amsmath]{revtex4}
\usepackage[dvips]{color,graphicx}
\usepackage{amsfonts}
\usepackage{amssymb}
\usepackage{latexsym}

\newcommand{\dalm}{\kern1pt\vbox{\hrule height 0.9pt\hbox{\vrule width
0.9pt\hskip 2.5pt\vbox{\vskip 5.5pt}\hskip 3pt\vrule width 0.3pt}\hrule height
0.3pt}\kern1pt}

\newcommand{\lw}[1]{\smash{\lower2.ex\hbox{#1}}}

\begin{document}

%\thispagestyle{empty}

%<<<<<<<<<<<<< TITLE >>>>>>>>>>>>>>>%
\title{On the measure of spacetime and gravity}
%<<<<<<<<<<<<< AUTHOR >>>>>>>>>>>>>>>%
\author{Naresh Dadhich\thanks{Electronic address:nkd@iucaa.ernet.in}}
\email{nkd@iucaa.ernet.in}
%<<<<<<<<<<<<< ADDRESS >>>>>>>>>>>>>>>%
\affiliation{Inter-University Centre for Astronomy \& Astrophysics, Post Bag 4, Pune~411~007, India\\}
\date{\today}

%======================================%
%<<<<<<<<<<<<< ABSTRACT >>>>>>>>>>>>>>>%
%======================================%
\begin{abstract} 
By following the general guiding principle that nothing should be prescribed or imposed on the universal entity, spacetime, we establish that it is the homogeneity (by which we mean homogeneity and isotropy of space and homogeneity of time) that requires not only a universally constant invariant velocity but also an invariant length given by its constant curvature, $\Lambda$ and spacetime is completely free of dynamics. Thus $c$ and $\Lambda$ are the only two true constants of the spacetime structure and no other physical constant could claim this degree of fundamentalness. When matter is introduced, the spacetime becomes inhomogeneous and dynamic, and its curvature then determines by the Bianchi differential identity the equation of motion for the Einstein gravity. The  homogeneity thus demands that the natural state of \textit{free spacetime} is of constant curvature and the cosmological constant thus emerges as a clear prediction which seems to be borne out by the observations of accelerating expansion of the Universe. However it has no relation to the vacuum energy and it could be envisioned that in terms of the Planck area the Universe measures $10^{120}$ units! \\
\end{abstract}


\maketitle
%\section{Introducation}

In the Newtonian mechanics, \textit{free} space in the absence of all forces is characterized by homogeneity and isotropy of space and homogeneity of time and particles in it follow its geodesics - straight lines. It is entirely characterized by these symmetry properties which is a dynamical identification of free space. Let us bring forth the full meaning of these properties in an imaginative way. Homogeneity in space would imply that the interchange of $x$ and $y$ coordinates would make no difference. Since time is also homogeneous, like the interchange of $x$ and $y$, the interchange of $x$ and $t$ should also be perfectly legitimate and acceptable. However the problem is that their dimensions don't match, we have to make them match so as to adhere to the general property of homogeneity of space and time. The only way the dimensions could be matched is by postulating the existence of a universally constant velocity, c, so then $x$ and $ct$ could be interchanged. It may be noted that this is all dictated by the sheer force of the consistency of a principle \cite{dad1}. The velocity of light thus enters into the picture and binds space and time into spacetime which harbours the Einsteinian mechanics, special relativity (SR). \\

The general principle we would like to adhere all through this discourse is that nothing should be prescribed or imposed on the universal entity, spacetime. The velocity of light binds space and time into spacetime, now the question arises,  what is its geometry? That is, what geometry should a homogeneous spacetime (here and in what follows by homogeneous we will always mean homogeneous and isotropic in space and homogeneous in time) which is free of all matter/energy in any form, have ? Since spacetime is homogeneous, its curvature must also be homogeneous which means it should be covariently constant, $\nabla_e{}R_{abcd} = 0$. The only thing that is constant relative to covariant derivative is the metric, hence homogeneous curvature is given by $R_{abcd} = \Lambda (g_{ac}g_{bd} - g_{ad}g_{bc})$, where $\Lambda$ is a constant, the measure of curvature of spacetime. Thus the geometry of homogeneous spacetime which is the analogue of Newtonian free space is in general curved and has constant curvature. Of course it would include the zero curvature flat spacetime as a particular case. In SR, geometry of spacetime was assumed to be flat described by the Minkowski metric while the dynamical characterization (homogeneity) of \textit{free} spacetime would have determined it to have constant curvature and not necessarily zero curvature. Minkowski geometry was therefore a prescription imposed on spacetime, it is not the natural and general geometry of free spacetime. What it means is that without any imposition from outside the \textit{free spacetime} has homogeneous (constant) curvature with $\Lambda$ as its measure. It would be described by dS/AdS metric depending upon the curvature being positive or negative and when it is zero, it is the flat Minkowski. \\

It is important to note that it is the property of homogeneity that first dictates the existence of the invariant velocity and now also the existence of an invariant length through the constant curvature of spacetime. Thus $\Lambda$ enters into the spacetime structure at the same footing as the velocity of light. It signifies spacetime's ability to curve as a basic innate property. As $c$ binds space and time into a $4$-dimensional spacetime and now $\Lambda$ curves it to give it a kind of a measure of its size through constant curvature. Thus these two constants get interwoven in the structure of spacetime and no other constant could claim this degree of fundamentalness. The basic property of homogeneity only demands existence of invariant velocity and length, it cannot say what should be their value? The invariant velocity is given by the Maxwell electrodynamics and it gets identified with the velocity of light, what then gives $\Lambda$? What we have seen so far is that the geometry of free spacetime should have constant curvature but what should be its measure cannot be determined unless spacetime is injected with some physical effects by introduction of matter/energy in some form or the other. It would then be determined by the matter content in the Universe. Ultimately it is for the observations to determine its value, however the accelerating expansion of the Universe seems to measure its value for the first time \cite{accel}. \\

%to define geometry of dynamics free state of spacetime. Like the velocity of light it now becomes the part of spacetime structure signifying the the spacetime's ability to curve. It is the introduction of matter which will make spacetime inhomogeneous and thereby dynamic. It   of This is a profound new realization that the spacetime (Universe) is not infinite but finite is what is implied by homogeneity without reference to matter or gravity. It is envisaged that the matter content in the Universe would determine its size by fixing the value of $\Lambda$. It may be noted that this is a prediction that naturally follows from this viewpoint.  It is ultimately for the observations to decide whether the spacetime has finite size or i\end{thebibliography}

%nfinite, that is its radius of curvature is finite or infinite. It is remarkable that the observations of the accelerating expansion of the Universe \cite{accel} are wonderfully consistent with the so called cosmological constant, $\Lambda$ implying the finite size. Thus the spacetime seems indeed to be finite bearing out the prediction and there is no need to invoke exotic dark energy for reconciling with the accelerated expansion of the Universe.  \\

It is clear from the above discussion that the general property of homogeneity demands not only an invariant velocity binding space and time into spacetime but also an invariant length \cite{dad1,pereira} given by the constant curvature of the  homogeneous \textit{free spacetime}. Thus the free spacetime is therefore homogeneously curved and free particles will now follow its geodesics - the motion under no forces. This is the state analogous to the one characterized by Newton's First law. It harbours no dynamics because it is homogeneous. Here there should be de Sitter relativity (dS-R) \cite{pereira} for $\Lambda > 0$ rather than SR of the Minkowski metric. Locally of course it is always Minkowski with the usual SR in the tangent plane. It should be noted that homogeneous spacetime described by dS/AdS metric is completely free of all dynamics including gravity. Though all the text books in GR state that the absence of gravity is characterized by vanishing of the Riemann curvature, we would however like to emphasize that it is instead characterized by the Riemann curvature being covariantly constant and not necessarily zero. It is the introduction of matter that would now make spacetime curvature inhomogeneous and thereby rendering it dynamic. The equation of motion governing this dynamics must also follow from the geometric properties of the curvature. The Bianchi identity satisfied by the Riemann curvature does indeed lead to an equation that interestingly describes the Einsteinian gravity. Gravity is thus driven by inhomogeneity in spacetime curvature. It is remarkable to note that gravity got automatically included in spacetime curvature because of its universal character of linkage to all that physically exist including zero mass particles \cite{dad2}. Since the dynamics of inhomogeneous spacetime must include homogeneous 
spacetime as a limit defining the zero of gravity, hence the equation must admit the constant curvature spacetime as the matter free solution.\\

Since gravitational dynamics is described by spacetime curvature, there is no freedom to prescribe a law of gravity, it should all follow from the Bianchi identity, $\nabla_[e{}R_{ab]cd} = 0$, satisfied by the Riemann curvature \cite{dad2}.  And it does indeed happen that the trace of the identity would give the divergence free symmetric Einstein tensor, $\nabla_b{}G_a{}^b = 0$, and consequently follows the equation,  
\begin{equation}
G_{ab} = \kappa T_{ab} - \Lambda g_{ab}, ~~~ \nabla_a{}T^{a}{}_b = 0  
\end{equation} 
where $G_{ab} = R_{ab} - \frac{1}{2}Rg_{ab}$ and $\Lambda g_{ab}$ is a constant relative to the covariant derivative. This becomes the equation for the Einsteinian  gravity when we identify $T_{ab}$ with the energy-momentum distribution - a universal property which is shared by all that physically exist (universal source for the universal force). Note that here $\Lambda$ enters the equation on the same footing as $T_{ab}$ and hence cannot simply be wiped out at one's whims and fancy without proper physical explanation and justification. More importantly the equation must admit homogeneous spacetime as the matter free solution and $\Lambda$ is its characterization. Had Einstein followed this natural geometric route, it won't have been then an addition as an after thought for obtaining the static cosmological solution \cite{eins} but would have rather emerged as a new constant of the Einsteinian gravity and spacetime? Further after the discovery of non-static expanding solution \cite{fried} and realizing the repulsive effect of $\Lambda$, he would have perhaps made a profound prediction though much before its time that the Universe may suffer accelerated expansion some time in future \cite{lambda}. It would have then been most remarkable when this would have actually been borne out by the observations \cite{accel}. \\ 

This derivation is based entirely on the differential geometric property of the Riemann curvature. The question arises, is it always possible to do so even when higher order terms in curvature are included? The answer is yes, it has been shown that the trace of the Bianchi derivative of a homogeneous polynomial in Riemann curvature with some properties similarly yields an analogue of the Einstein tensor corresponding to each term in the Lovelock polynomial \cite{dad3}. In fact it happens only for the Lovelock polynomial action and hence it defines a new characterization of the Lovelock gravity. \\

A field equation has one free constant which is determined by experimentally measuring the strength of the field. The above equation for gravitational field has instead two constants, one of which measures the strength of the force and is identified with the Newtonian constant, $\kappa = -8\pi G/c^2$. Why does it have the additional constant $\Lambda$? The feature which is different for gravity from all other forces is that it has no given fixed background spacetime and hence the equation must also contain the information of the state of the reference spacetime against which gravitational field is measured. That is what the new constant, $\Lambda$ refers to and it is a homogeneous spacetime of constant curvature as the background reference. Since homogeneous spacetime should be the natural reference for inhomogeneous spacetime, it should also be a natural reference for gravitational field which is described by inhomogeneous spacetime. \\

The so called cosmological constant, $\Lambda$, as we all know, had a very chequered history mainly because the way Einstein introduced it as an after thought for having a static model of the Universe \cite{eins}. Once the non-static solution for the Universe was found \cite{fried}, its presence was no longer needed for the purpose for which it was introduced. Thus remained the situation until the vacuum energy arising out of quantum vacuum fluctuations was considered \cite{zeldo} which against flat spacetime background has the stress tensor of the same form as $\Lambda g_{ab}$. But then its value was off the Planck length by staggering 120 orders of magnitude. Once again, it went into deep slumber to rise again with the renewed vigour to describe the observed acceleration of the expansion of the Universe \cite{accel}. Notwithstanding the very large number of dark energy models of enormous variety and vital statistics as well as other models involving modified  gravity and inhomogeneity, the observational data agrees with $\Lambda$ admirably well. We would strongly like to argue that though the vacuum energy must, like the gravitational field energy, gravitate but in a subtler way than writing a stress tensor on the right of the equation. \\

The Einstein gravity is indeed self interactive yet the gravitational potential is however given in $g_{tt} = 1 + 2\phi$ by $\phi = -M/r$, which obviously includes no self interactive contribution. How is it then taken care of? It is in fact done through the null energy condition implying $g_{tt}g_{rr} = -1$, and that  means $3$-space must be curved. It is the curvature of $3$-space which accounts for the gravitational field energy contribution leaving the Newtonian potential intact \cite{dad4}. Intuitively we can say Einstein is therefore Newton with $3$-space curved. Note that the gravitational field energy is not a primary source as it has no independent existence of its own because gravitational field is created by matter. This is why it cannot sit through a stress tensor alongside the primary matter source ($T_{ab}$) in the equation (it is a different matter that one cannot in principle write a covariant stress tensor for it). Its contribution can only be incorporated by curving $3$-space. That is by enlarging the spacetime framework and not by adding a term in the equation. The situation with the vacuum energy is exactly the same as it is also created by matter and has no independent existence of its own. Like the gravitational field energy, it must not gravitate through a stress tensor on the right of the equation but should ask for the enlargement of the spacetime framework for its inclusion. Since the vacuum energy is a quantum phenomenon, enlargement of the framework would emerge only when we have quantum theory of gravity. So long as that doesn't happen, we cannot truly understand gravitational interaction of the vacuum energy. For a clear and correct understanding of its gravitational interaction, there is therefore no shortcut but to wait for quantum gravity. \\

There is also a case made for tracefree gravitational equation in which only the  tracefree matter gravitates \cite{wpe}. Since the vacuum energy has non-zero trace,  hence it cannot gravitate and then $\Lambda$ appears in the derived equation as an integration constant. This is against the basic property of the universality of gravity which requires its linkage to all that physically exist. All kinds of energy must gravitate, however mode of interaction could be different. The main concern here is to make $\Lambda$ free from the vacuum energy and so it is not measured against the Planck length. However there arises a basic problem in treating the vacuum energy as a fluid. Its stress tensor implies the equation of state, $\rho + p = 0$. In the fluid equation of motion, $\rho + p$ defines the inertial density similar to the inertial mass in the particle equation of motion. Clearly the equation becomes untenable whenever inertial mass or density vanishes. This signals the need for a new theory for inclusion of such a situation. Recall that the zero mass particle required an invariant velocity for its description and thereby asking for a new mechanics - special relativity. Note that what was required for inclusion of zero mass particle was the enlargement of the framework signified by merger of space and time into spacetime. And so should be the case for inclusion of the zero inertial density. It can therefore be not included by writing a stress tensor on the right but we have to seek enlargement of the spacetime framework. What could that be? Since it arises from the quantum fluctuations of vacuum, it would require a discrete quantum structure for space to suffer fluctuations. The enlarged framework would be provided by the way this basic feature of space is incorporated in a new theory of quantum spacetime/gravity. Then in the enlarged framework it would be automatically taken care of. A general principle is that if something (like $m=0$ particle) is not admitted in the existing framework, the framework is enlarged in such a way that it gets automatically incorporated \cite{dad5} and there comes up a new theory encompassing the existing one. This is precisely what should happen for the vacuum energy when we have the quantum theory of gravity. \\

Let us reemphasize that $\Lambda$ stands all of its own on the shoulders of the basic spacetime property of homogeneity without reference to matter, gravity or  cosmology. It describes the geometry of spacetime completely free of all matter and dynamics but there is no way to find its value in this free state of spacetime. What it signifies is the fact that curvature is the innate property of spacetime even in the absence of matter. It therefore provides a right framework of curved spacetime for the description of gravity and also serves as the reference (homogeneous) for the measure of presence of matter and gravity (inhomogeneous). The introduction of matter thus does not produce any discontinuity in spacetime structure from flat to curved instead it is just a continuous transition from homogeneous to inhomogeneous. That is why $\Lambda$ is indeed a true constant of spacetime structure and thereby of the Einstein gravity \cite{rovelli, dad6}. Since it is, as argued above, not related to the vacuum energy and hence there is no reason for it to be slated against the Planck length. It can now have any value like any other physical constant. Since it refers to the spacetime which means the Universe as a whole, it would be the cosmology that would fix its value. In the Friedmann model of the (spatial) homogeneous and isotropic FRW cosmology \cite{fried}, it is determined in terms of the energy density and the Hubble expansion parameter. Interestingly it turns out that $\Lambda$ determined by the present values of these parameters agrees wonderfully well with the accelerating expansion observations \cite{accel}. As the value of invariant velocity was fixed by the electrodynamics as the velocity of light, similarly $\Lambda$ is fixed by the cosmology in terms of the matter content in the Universe and the Hubble parameter. The critical value for $\Lambda$ is given by  $\Lambda = \Lambda_c = 4\pi G \rho_0/c^2$, where $\rho_0$ is the bounce value of the density. It then turns out that for the spherical ($k=1$) case it would be like the Friedmann recollapsing model for $\Lambda < \Lambda_c$ while for $\Lambda > \Lambda_c$, it would expand to ultimate dispersion resembling the de Sitter \cite{jvn}. In the absence of $\Lambda$, it is the curvature parameter $k$ that separates the recollapsing and ever expanding cases. Since $\Lambda > 0$ has dispersive effect, it has to be bounded from above so that it does not dominate over the density for the Universe to recollapse. \\

The enlargement of framework is always required for incorporation of some new physical feature or entity like the zero mass particle which is not admitted in the existing framework and then the framework is enlarged in such a way that it gets  included \cite{dad5}. In the case of zero mass particle, a universal constant velocity was required which bound together space and time into spacetime which provided the enlarged framework. Since the vacuum energy is a quantum phenomenon, the enlargement has to be sought involving quantum properties of spacetime. Where does and in what form should one seek the enlargement of framework? Since we are asking for a quantum theory of spacetime, it is natural to expect that at deep down spacetime may also have discrete quantum structure like matter. Would the usual quantum principle and techniques be appropriate to handle it when spacetime which provides the background itself has quantum structure or a new framework would be needed? What could be its building blocks describing the microstructure? This is what the loop quantum gravity adherents are attempting to probe \cite{abhay}. Further are there any unexplored properties of spacetime and gravity that could be invoked? One could be that spacetime may not retain its commutative character at the micro level \cite{noncom} and the other could be that gravity may not entirely remain confined to the usual $4$ dimensions. It may leak into higher dimensions as envisaged in the braneworld gravity \cite{rs}.  For higher dimensions, apart from the string theory which naturally lives in there, there are quite strong classical motivations as well \cite{dad5,dad3}. As and when we have the quantum theory of gravity, we believe that the enlarged framework would take care of the vacuum energy in the similar fashion as gravitational field energy is taken care of through the space curvature in general relativity. \\

It would be illuminating to discuss one of the interesting arguments for the extra dimensions \cite{dad5,dad3}. It is based on the general principle that the total charge for a classical field must always be zero globally. This is true for the  electromagnetic field, how about gravity? The charge for gravity is the energy momentum that is always positive, how could that be neutralized? The only way it could be neutralized is that the gravitational field the matter creates itself should have charge of opposite polarity. That means the gravitational interaction energy must be negative and that is why the field has always to be attractive. Note that gravity is attractive to make the total charge zero. The negative charge is however not localizable as it is spread all over the space. When it is integrated over the whole space for a mass point, it would perfectly balance the positive mass.  This is what was rigorously established in the famous ADM calculation \cite{adm}.  Consider a neighbourhood of radius $R$ around a mass point. In this region, the total charge is not zero and there is overdominance of positive charge as the negative charge lying in the field outside $R$ has been cut off. Whenever the charge is non zero on any surface (like for an electrically charged sphere, the field propagates off the sphere) the field must go off it. This means gravity must go off the $3$-brane ($4$-spacetime) in the extra dimension but as it leaks out, its past light cone would encompass more and more region outside $R$ and thereby more and  more the negative charge. That is it leaks off the brane with diminishing field  strength and hence it does not penetrate deep enough. This is the picture quite similar to the Randall-Sundrum braneworld gravity model \cite{rs}. If the matter fields remain confined to the $3$-brane \cite{dim} and only gravity leaks into extra dimensions but not deep enough, then extra dimensions effectively become small (for probing depth of dimension, we need some physical probe that goes there). This is an intuitively very appealing and enlightening classical consideration why gravity cannot remain confined to $4$ dimension and at the same time why extra dimension cannot be large?

In conclusion, we note the main points of the discourse. It is the inhomogeneity of the spacetime curvature that drives the gravitational dynamics while homogeneity signified by $\Lambda$ provides the zero reference for gravity. It thus becomes the part of the spacetime structure as much as the velocity of light. Above all the natural state of spacetime is always curved, when it is free it is homogeneously curved and the introduction of matter makes it inhomogeneous and then it harbours the Einstein gravitational dynamics. Thus introduction of matter does not imply any break in the spacetime structure from flat to curved as is required in the standard  picture. The spacetime like anything that bends (matter) should have discrete  quantum micro-structure. It is required for it to curve so that it can describe gravitational dynamics. There are also indications of the converse being true. That is a constant curvature de Sitter spacetime can be shown to emerge from the classical sequential growth of causal sets \cite{ride}. The gravitational interaction of the vacuum energy, like that of the gravitational field energy, cannot be incorporated by writing a stress tensor on the right of the Einstein equation but instead has to be done by enlarging the spacetime framework which would unfortunately unfold only when the quantum theory of gravity is discovered. Until then there is no alternative but to wait. With $\Lambda$ being completely free of the vacuum energy, it can have any value that the FRW cosmology fixes for it in terms of the energy density and the Hubble expansion parameter. Its value as determined by the present values of these parameters agrees wonderfully well with the accelerating expansion observations \cite{accel}. Further it is conceivable that had Einstein followed this chain of thought, he could have perhaps predicted the accelerated expansion some time in future. If that were so, it would have been one of the most profound predictions of all times and we would have once again saluted his genius when it would have been verified by the observations. This is the simplest and clearest explanation of the observations without any need for the exotic dark energy. Lastly let us turn an embarrassment into a virtue by turning the argument on its head, it could as well be envisioned that in terms of the Planck area the Universe measures $10^{120}$ units!

--------------


\begin{thebibliography}{9}
%----------------------------------------
\bibitem{dad1} N. Dadhich, Physics News {\bf 39}, 20 (2009), arxiv:1003.2359.

\bibitem{accel} S. Perlmutter et.al, Astrophys. J. {\bf 483}, 565 (1997); Nature, {\bf 391}, 51 (1998).

\bibitem{pereira} R. Aldrovandi, J. P. Beltran Almeida and J. G. Pereira, 
Class.Quant.Grav. {\bf 24}, 1385 (2007).

\bibitem{dad2} N. Dadhich, Subtle is the gravity, gr-qc/0102009.

\bibitem{eins} A. Einstein, Cosmological considerations in the general theory of relativity, Sitzungsber. Preuss. Akad. Wiss. Berlin (Math Phys.) 142-152 (1917).

\bibitem{fried} A. Friedmann, Z. Phys. {\bf 10}, 377 (1924).

\bibitem{lambda} N. Dadhich, On the enigmatic $\Lambda$ - a true constant of spacetime, arxiv:1006.1552v2.

\bibitem{dad3} N. Dadhich, Pramana, {\bf 74}, 875 (2010), arxiv:0802.3034.

\bibitem{zeldo}  Y. B. Zeldovich, Usp. Fiz. Nauk, {\bf 89}, 647 (1966).

\bibitem{dad4} N. Dadhich, On the Schwarzschild field, gr-qc/9704068.

\bibitem{wpe} S Weinberg, Rev. Mod. Phys. {\bf 61},.1 (1989); T. Padmanabhan, Gen. Rel. Grav. {\bf 40}, 529 (2008); G. Ellis, J. Murugan and H. van Elst, The gravitational effect of the vacuum, arxiv:1008.1196,v1.

\bibitem{dad5} N. Dadhich, Universalization as a physical guiding principle, gr-qc/0311028.

\bibitem{rovelli} Bianchi and C. Rovelli, Why all these prejudices against a constant?, arxiv:1002.3966v3.

\bibitem{dad6} N. Dadhich, On the enigmatic $\Lambda$ - the true constant of spacetime, arxiv:1006.1552.

\bibitem{jvn} J. V. Narlikar, Introduction to Cosmology (Cambridge Univ. Press, 1993).

\bibitem{abhay} A. Ashtekar, Quantum spacetimes: Beyond the continuum of Minkowski and Einstein, arxiv:0810.0514.

\bibitem{adm} R. Arnowitt, S. Deser, C. W. Misner, in Gravitation: An Introduction to Current Research, ed. L. Witten (John Wiley, 1962) p.227.

\bibitem{noncom} N. Franco, Survey of gravity in non-commutative geometry, arxiv:0904.4456.

\bibitem{rs} L. Randall and R. Sundrum, Phys. Rev. Lett. {\bf 83}, 4690 (1999).

\bibitem{dim} N. Dadhich, Why do we live in four dimension?, arxiv:0902.0205. 

\bibitem{ride} M Ahmed and D Rideout, Phys. Rev. {\bf D81}, 083528 (2010), arxiv:0909.4771.


\end{thebibliography}
\end{document}